\documentclass[twocolumn,aps,preprintnumbers,amsmath,amssymb]{revtex4}
\usepackage{graphics}
\usepackage{graphicx}
\usepackage{dcolumn}
\usepackage{bm}
\usepackage{epsfig} 
\usepackage{psfrag} 

\usepackage[dvips]{color}

\begin{document}


\title{\bf An interaction model on random networks: from social to airports networks.}

\author{Philippe Curty}

\affiliation{Scimetrica Research, www.scimetrica.ch, 3007 Bern, Switzerland}

\begin{abstract}
\bf \sf
Social movements, neurons in the brain or even industrial suppliers are best described by
 agents evolving on networks  with basic interaction rules. In these real systems, the connectivity
  between agents corresponds to the a critical state of the system related to the noise of the system. The new idea is that
  connectivity adjusts itself because of two opposite tendencies: on the one hand informations percolation is
  better when the network connectivity is small but all agents have rapidely the same state and the dynamics stops.
  On the other hand, when agents  have a large connectivity, the state of a node (opinion of a person, state of a neuron, ...)  tends to freeze: agents
  find always a minority among their neighbours to support their state.
The  model introduced here captures this essential feature showing a clear transition between the two tendencies at some
 critical connectivity. Depending on the noise, the dynamics of the system can only take place
 at a precise critical connectivity since, away from this critical point, the system remains in a static phase.
 When the noise is very small, the critical connectivity becomes very large, and highly connected
  networks are obtained like the airports network and the Internet.
 This model may be used as a starting point for understanding the evolution of agents living on networks.

\end{abstract}


\maketitle

 What are the necessary conditions allowing a consensus among an assembly of voters?
 Why, sometimes, do all people share the same opinion in a short time?
For example, we consider a school class of students who want to make an important travel, and have to
 choose between two destinations: Alaska or Rio de Janeiro. Every student has
 naturally a strong preference for one of the destinations. If students are isolated then nobody changes
  his opinion, and no consensus shall be found.
When students take advice from one or two friends in the class, they may change their opinion quickly
 and a consensus will be achieved. Now if each student has many friends, he will always find a small
  group among his friends that shares his opinion and support his choice. Hence students will keep their
   opinion and no consensus is found although every student is connected to a large number of friends.
In a neural network, the number of connections among neurons, dendritic tree, plays an important role
 like a social network. What is the critical number of incoming connections on a neuron? In the following
  article, we will see how to reproduce these behaviours using the language of opinion dynamics, and
  analyse the consequences on various fields like supply networks of firms or the airports networks.

There exist already several models where opinion dynamics of a
community of agents is simulated. Among them are the voter model \cite{clifford1973}:
a 2-states spin is selected at random and it adopts the opinion of a
randomly-chosen neighbour. This step is repeated until the system reaches
consensus. Each agent has zero self confidence since he merely adopts
the state of one of his neighbours.
A similar model due to K. Sznajd-Weron and J. Sznajd  \cite{sznajd2000} was designed to
explain certain features of opinion dynamics resumed by the slogan  ''United we stand,
divided we fal''. This  leads to a dynamics, in that individuals placed on a
lattice can choose between two opinions, and in each update a pair of
neighbours sharing common opinion persuade their neighbours to join their
opinion. It is equivalent to the voter model as shown in ref.  \cite{behera2003}.
The local majority rule model described in ref. \cite{huepe2002,krapivsky2003} considers groups of
 agents where members adopt the opinion of the local majority.

It is commonly believed that correlated behaviour in financial markets or large opinion changes in
human society is due to informations that are shared by everyone at the same time. For example in a
 financial market, herding can be produced by the central market price, or by a rumor that
propagates rapidly among traders. Informations are collected from the action of each traders and is
reflected in the price of stocks. However sometimes opinions changes occurs only because of
local interaction among agents as we shall see later.

Some models describe the consequence of ''herding'' behaviour \cite{cont2000,eguiluz2003,curty2006} in financial markets.
However, in this class of models, herding behaviour is controlled by an external parameter. The question answered by this class of models is
 related to the consequence of herding on distribution of financial
price data. In our case we are interested at finding the source of herding,
why people follow sometimes the same trend in a short time although they seem to act independently.

The evolution of a genetic network has been described by S. Kauffman \cite{kauffman1993} using andom boolan networks (RNB). In RBN  the state of a node is determined by the state of $G$ neighbours depending a random lookup table. On the contrary, the model introduced here uses the same rule for all agents: the state of an agent changes only if all $G$ neighbours have the opposite state. A temperature or noise allows agents to change their state by themselves. This rule is simpler, and maybe more general, compared to the RBN model since all the randomness is resumed in a single noise parameter $r$ rather than in complex lookup tables.

The aim of the present work is to study the dynamics of a
 ensemble of agents connected on a random network, and to determine what are the conditions to
 have consensus or disorder under a fundamental interaction rule.

\noindent
{\bf State - } Each agent has an opinion (state)  $O \in \{-1, 0, +1\}$ with three possible values that can represent
\{vote A, do not vote, vote B\} for a vote, or  \{buy, wait, sell\} in a financial market.
In real life, an agent can be a single person or a group of people that
share the same opinion.
The formation of opinion is determined by the confrontation, i.e. summation, of the opinion of the
agent with the opinion of each advisor. Contrary to the voter model or
the majority model \cite{krapivsky2003}, here agents have a the same self-confidence or strength as each of their advisors,
provided they have an opinion +1 or -1.

\noindent
{\bf Algorithm (fixed connectivity $G$) - } Now we consider a community of $N$ agents where each agent
$i$ has an opinion $O_i(t) \in  \{ -1, 0, +1\}$ at time $t$.  Each agent can be either an advisor or being
 advised by other agents.
In each update, we sum the opinion of an agent picked at random  with each opinion of $G$ advisors
 chosen at random among all agents.
The sign of this sum represents his new opinion. More explicitly, at each time step $t$:

\begin{enumerate}
\item  An agent $i$ is selected at random.
\item  The new opinion $O_i(t+1)$ is the sign of the sum of the opinion of the agent
$i$ with each opinion  $O_{i_k}(t)$ of a random advisors group $A_i= \{i_1, ... , i_G\}$:
\begin{equation}
O_i(t+1) = \mbox{sign} \left( \sum_{k=1}^G [O_{i}(t) + O_{i_k}(t)] \right)
%
\label{recursion}
\end{equation}
where $\mbox{sign}(0)=0$.
\item Instead of point 2., with probability $r$, the new opinion  $O_i(t+1)$ is  +1 or -1 taken at random.\\
\end{enumerate}

The recursion relation (\ref{recursion}) means that agents change their opinion only if all advisors
 have an opposite opinion (unanimity rule). The algorithm is completely deterministic when no
 random opinion is introduce during the simulation, i.e. if $r=0$. The situation $r>0$ is more realistic
 since it is reasonable to assume that people change their opinion sometimes at random.

The two-step change of opinion is realistic since people may have state where they are not active, but this state is not essential to the dynamics of the system. A one-step change would essentialy lead to the same results.

The key parameter is the number of advisors per agent or connectivity $G$.
For $G=1$ agent merely change the opinion when the advisor has an opposite opinion. In this case
 the dynamics is similar to the voter model \cite{clifford1973}. For $G=2$, i.e. one agent and two advisors,
the algorithm is equivalent to the majority model \cite{krapivsky2003}: if advisors have the same
 opinion, they form the majority of the three agents (two advisors plus the advised agent).
 The two limiting cases are then:

\noindent
{\it  Small connectivity (hierarchical society)}: each agent follows a small number of advisors
and opinions can change easily.
Starting from a random configuration, opinions or informations scatter rapidly through the network resulting
in a rapid consensus. A long range order appear due to strong correlations between advisors and agents.
The majority and voter models belong to this category, and are therefore opinion dynamics
 with weak self-confidence and rapid decision making.

\noindent
{\it Large connectivity (complex society)}:
Agents tend to keep their own opinion because the probability that all advisors have the same and opposite opinion is small. Informations cannot be transmitted through the network, and eventually no long  range order
or consensus can emerge: opinions are essentially random.
The diversity or complexity of a community increases with its size.
The connectivity $G$ of an agent is proportional to his self-confidence: agents with many advisors have large
self-confidence since they keep their opinion when they find at least one advisor sharing their opinion.

\begin{figure}[h]
\centering
\includegraphics*[viewport=0 0 720 480,angle=0,width=8.5cm]{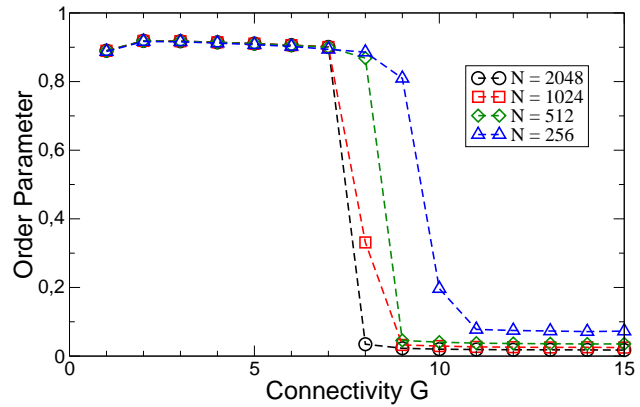}
\vspace{-0.2cm}
\caption{ \label{fig-opinion-average-fix}
Average absolute opinion (order parameter) $\tilde O(G)$ for connectivities $G \in\{1,2,...\}$
and different system sizes. The noise is $r=0.05$. The transition between consensus and disorder
is located near $G_c \approx 8$. The variance (not shown)  is always 0 except for a non-zero order parameter at the transition (Here N=1024, G=8 and N=256, G=10).
}
\end{figure}

\noindent
{\bf Results - } Computer simulations have been done using different values of the noise $r$ and
different group sizes. Statistics are done over 10 to 20 runs where each run has
$10^4$ updates per agent.
As shown in figure  \ref{fig-opinion-average-fix} for $r=0.05$, the order parameter (average absolute opinion)
\begin{equation}
{\tilde O} :=  {1 \over N} \langle \  | \sum_{i} O_{i}(t) | \ \rangle_t,
\end{equation}
which is averaged over time $t$, has a breakdown at a critical group size $G_c(r)$ marking a
clear separation between two different regimes.
 For $G>G_c$, there is no global coordination.
 Opinions are essentially random and $\tilde O=0$. At $G=G_c$, opinions oscillate between
 ordered and random states.  For $G<G_c$, opinions are correlated and rapidly evolve
 either to the average consensus -1 or +1 depending on initial conditions: $\tilde O > 0$.
 A similar noise driven phase transition has been found for the majority model in ref.
 \cite{huepe2002}, and it corresponds in our model to the case $G=2$.

When all links between agents are reciprocal no phase transition occurs: reciprocity
reduce the transmission of informations through the network.

\begin{figure}[h]
\begin{center}  \setlength\unitlength{1cm}
\begin{picture}(8.5,5)
\put(0,0){\resizebox{8.5cm}{!}{\includegraphics{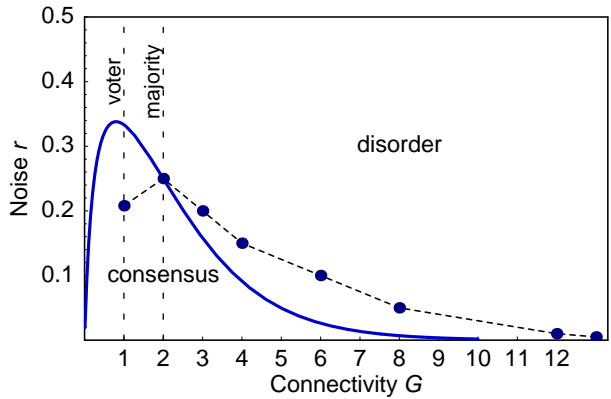}}}
\end{picture}
\end{center}
\vspace{-0.5cm}
\caption{ \label{fig-phasediagram-static}
Phase diagram in the $\{r,G\}$ plane showing two different regimes: consensus and disorder.
 The thick line is the analytical solution $r_c(G)$ from equation (\protect{\ref{rc}}). Points
are results from computer simulations.}
\end{figure}

\noindent
{\bf Analytical Approach - }
The critical point $G_c$, found in simulations, separating the correlated phase and the random phase
can be derived analytically. To do that, we consider a system with
a infinite number of agents, and we neglect effects of loops.
The connectivity $G$ of an agent is then equal in average to the number of agents that an agent
advises, i.e. $G = G_{out}$.

Now we look for the probability $P$ that an agent changes his opinion.
Only opinions that are -1 or +1 are taken into account since 0 opinions disappear quickly.
If we have $n$ agents with opinion +1, then   $x=n/N$  is the probability of finding an agent with opinion +1.
If the noise is zero, the probability $P_{G}$ that an agent $i$ changes his opinion from -1 to +1 or from +1 to -1 is given by the probability
to find the agent $i$ with opinion $-1$ and $G$ advisors with opinions $+1$ plus the corresponding
 probability to find the agent $i$ with opinion $+1$ and $G$ advisors with opinions $-1$:
\begin{eqnarray}
P_{G} = {G \over 1+G} \ \left[ x (1-x)^G + (1-x) x^G\right]
\end{eqnarray}
where the factor ${G \over 1+G}$ is introduced in order to take into account that empty groups, i.e. $G=0$,
induce no opinion change. If we add a noise $r$ with uniform distribution between 0 and 1, the total probability
$P$ of changing the opinion is 1/2 with probability $r$ plus $P_{G}$ with probability $1-r$:
\begin{eqnarray}
P =   {r \over 2} + (1-r) P_G.
\label{total-probability}
\end{eqnarray}
Consensus is reached when the probability of changing the opinion because of the advisors is larger than the probability of changing randomly the opinion:
\begin{eqnarray}
(1-r) P_G > r/2  \quad \Rightarrow  \quad \mathrm{consensus}.
 \label{criterion}
\end{eqnarray}
For a random configuration with $x=1/2$, the critical noise $r_c$, which separates
consensus and disorder, is determined by the condition
$(1-r_c) P_G = r_c/2$. This leads to:
\begin{eqnarray}
r_c = \frac{1}{1+2^{G-1} \left(1+G^{-1}\right)}.
\label{rc}
\end{eqnarray}
$r_c$ has a maximum at $G \approx 0.801$.

In figure \ref{fig-phasediagram-static}, the results of simulations are shown together with the analytical results from
equation (\ref{rc}) showing a qualitative good agreement. Note that the majority rule ($G=2$) leads to a larger consensus phase than $G=1$ (dynamics similar to the voter model).
The transition points for the infinite system are obtained by computing the intersection of reduced fourth order cumulants
 \cite{binder1981} for different system sizes.

\noindent
{\bf Dynamical Groups - } Until now, each agent had the same number of advisors. A drawback of this
static approach is that the number of advisors is a discrete quantity. Hence it is not possible to study the phase
transition as a function of a continuous parameter. Moreover a fixed number of advisors
is not very realistic because people usually have different numbers of advisors.
In order to get closer to reality and to study the transition with a continuous parameter,
advisors groups are now formed according to a probability $p$ of growing the group size by one advisor.
 The algorithm starts with empty groups, and at each time step $t$:\\
\noindent
1. An agent $i$ is selected at random.\\
2. With probability $p$, increase the group size by one advisor: $G_i \rightarrow G_i +1$.
Otherwise, i.e. with probability $1-p$,  perform point 2. of the static model, and remove one advisor:  $G_i \rightarrow G_i -1$.\\
3.  Instead of point 2., with probability $r$, a new opinion +1 or -1 is taken at random.\\

\begin{figure}[h]
\begin{center}
\includegraphics*[viewport=0 0 720 480,angle=0,width=8.5cm]{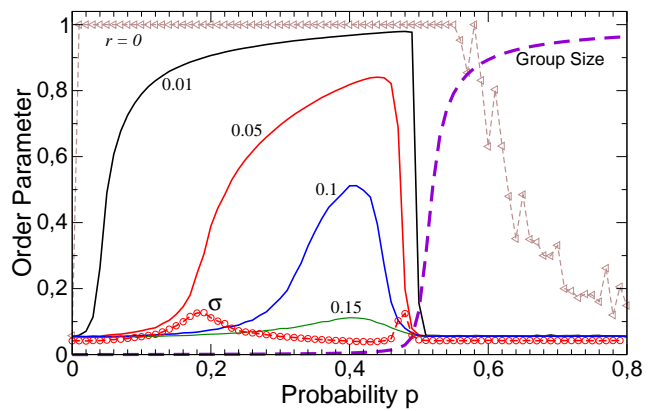}
\end{center}
\vspace{-0.4cm}
\caption{ \label{fig-order_parameter_dynamical}  Average absolute opinion (order parameter) $\tilde O$
of the dynamical model for $N=400$ agents and different value of noise $r$, and the corresponding
average relative group size $\langle G_i \rangle/N$. The circles show the standard deviation
 $\sigma$ of $\tilde O$ for $r=0.05$.}
\end{figure}

In figure \ref{fig-order_parameter_dynamical}, the order parameter $\tilde O$
is plotted for different values of the noise $r$ for $10^4$ steps and 20 runs. Following a line
defined by a constant order parameter: when $p$ is very small, most agents have zero advisor, and  informations cannot percolate
through the network ($\tilde O=0$). When $p$ increases,
there is a critical probability $p$ where agents have enough connections in order to establish
consensus in the entire network, and $\tilde O>0$.
 If we increase again $p$, a second transition occurs where groups of advisors are so large
  that agents cannot change their opinion anymore ($\tilde O=0$).
Simulations for $r=0$ are not conclusive since they suffer of a very slow dynamics.
The standard deviation  of $\tilde O$ is also reported and it exhibits maxima
at the two transition points.

\begin{figure}
\begin{center}  \setlength\unitlength{1cm}
\begin{picture}(8.5,5.5)
\put(0,0){\resizebox{8.5cm}{!}{\includegraphics{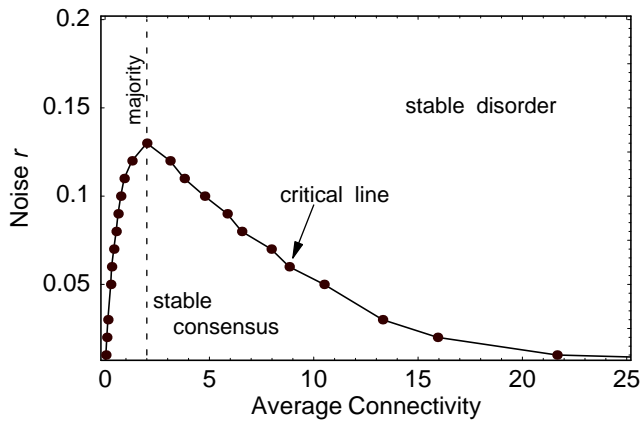}}}
\end{picture}
\end{center}
\vspace{-0.5cm}
\caption{ \label{fig-phasediagram-dynamic}
The critical connectivity (points) marks the transition between stable consensus and
stable disordered phases. The transition
coincides with the maximum of  the standard deviation or activity  $\sigma$  of the order parameter.
Computer simulations have been done for different noise $r$ and probability $p$. }
\end{figure}

The phase diagram of the dynamical model is shown in figure \ref{fig-phasediagram-dynamic}.
The transition points for the infinite system are obtained with fourth order cumulants for different system sizes.
We note that for small groups $G$, the transition occurs at a smaller noise $r$ compared with the static model
shown in figure  \ref{fig-phasediagram-static} whereas $r_c$ is larger for big groups.

 An interesting feature is that the majority rule $G=2$ is the less noise sensitive of all decision procedures.
 This is a hint showing that communities use in general the majority rule because it is the most
error tolerant system.

\noindent
{\bf A necessary condition of the evolution - } we consider now the class of systems evolving on random network and constrained by
the unanimity rule. These systems can be a social network, industrial supply networks or airports networks.
Considering a given error tolerance or noise, there is a unique average critical connectivity $\langle G_c \rangle$ where the system has
a phase transition between a consensus phase and a disordered phase. If the system decreases its connectivity, individuals or nodes
 can change their state according to their advisors resulting in a rapid consensus: then all nodes have the same state.
 This consensus phase is stable and does not
allow nodes to change their state anymore and no evolution is possible. When the system increases its connectivity, nodes tend to keep their states ending in
a disordered phase. This disordered phase remains frozen as long as the connectivity is large and nodes keep their state. In this static phase, the system cannot move from one configuration to another, and therefore no  evolution can take place.
The only point where this class of systems can change their state  is at the average critical connectivity $\langle G_c \rangle$. Depending on the particular noise of the system,  at $\langle G_c \rangle$, the agents of the systems change  all the time between consensus and disorder. Since the system is
in a critical and unstable state, a small perturbation in one node can then result in avalanche of changes in a large part of the system. This critical state is therefore the only region where the evolution of the system can take place.
Note that the evolution of the underlying random network itself is a related but different problem. This simple model is not a complete model of evolution because there is no selection rules or genetic evolution like in real systems.

\noindent
{\bf Social networks - } In a  real networks of people,  the probability
$p$ can be interpreted as a "social pressure" which forces people to be near an ideal number of advisor.
 Extreme opinions that propagate rapidly are dangerous for the stability of the social cohesion.
On the other side, people taking into account too many different sources of informations are unable to
change their opinion. Hence there is a critical number of advisors $\langle G_c\rangle$ that allows the system to change its state.
 If we define the "social activity" as the variance  $\sigma$ of the order parameter $\tilde O$, then $\sigma$ is maximum precisely at the transition
as shown in figure \ref{fig-order_parameter_dynamical}. We consider only the upper transition, and not
the percolation transition where $G_c$ is small.
This criticality is related to the concept self-organised criticality as introduced by P. Bak {\em et al}
in ref. \cite{bak1987}. In a real process of decision making, the main difficulty is to estimate the noise level that is present.
For example a noise level of $r =0.1$, which seems reasonable, leads to
$G_c \approx 5$. Of course the number of real contacts, either groups or single persons, of an
agent is larger than $G_c$ since not every contact is an advisor. The random network of the present model can be therefore different from the physical network since advisors may consist of group of persons (for example, the family).

\noindent
{\bf Supply networks - } A complex object like a car needs
several components produced by other firms. These firms transform material produced by other firms as well. Hence the
flux of products forms a network where the product can successfully reach its destination (+1) or not (-1).
Like in other networks, there is a critical connectivity in the firms network: it is more difficult to have
hundred of suppliers since coordination costs increases with the coordination number. On the other hand, having a few
 suppliers induces dependence and decreases flexibility. Therefore for each suppliers network, there is an critical number $G_c$ of
suppliers depending on the error tolerance $r$. These conclusions can be seen as a generalisation of empirical cost
studies for a single firm as a function of the number of suppliers (see \cite{bakos1993}).

\noindent
{\bf Networks with large connectivity - } a special case is obtained for zero noise, $r \rightarrow 0$: the critical connectivity tends to infinity,
i.e. $G_c \rightarrow \infty$ as shown by equation (\ref{rc}). The network never finds its equilibrium and, although a finite
 average connectivity can be calculated or measured, it does not correspond to the critical connectivity $G_c$ that goes to infinity.
 In the airports network, the noise level is very low ($r \rightarrow 0$): airports are only in function when
  all airplanes arrive at destination (unanimity rule). The routers in the Internet network are subject to zero error tolerance as well.
A router is operating only when all incoming packets are distributed without error. Hence, nodes of these networks
can accumulate connections as long as the noise $r$ is very low, and they may end to a scale free networks.

The growing process of scale free networks has been described in ref. \cite{barabasi1999}:  large nodes
have a bigger probability to have additional links. However, the Barabasi algorithm provides no explanation so far about the evolution of system living on scale free networks.\\

In conclusion, a general model of interactions on random networks is presented in the framework of dynamics of agents with advisors or neighbours.
A unanimity rule is introduced, and by varying the connectivity, one can tune the stability of agents. This model is as
generalisation of existing opinion models: for one advisor, a dynamics similar to the voter model is obtained, and two advisors
is equivalent to a majority rule.
When varying the connectivity $G$, a phase transition occurs at a certain value $G_c$ between a phase of
 consensus and a phase of disorder. $G_c$ depends
on the noise or error tolerance of the system and on the particular underlying network.
When agents have a few number of advisors, informations percolate through the network since agents change
their opinion frequently, and a consensus is found.
In the case of large advisor groups, agents tend to keep their opinion since large groups have less frequently the opposite opinion as the agent, and a little noise causes disorder.
Finally, it is shown that real social networks and industrial networks organise themselves
at the maximum of activity, precisely at the critical connectivity $G_c$ that sets a scale for the system
depending on the error tolerance. The airports network or the Internet, where error tolerance is very low, have a
  large critical connectivity tending to infinity.

A future issue is the role of the structure of the network and the possibility to have heterogenous noise in the system. Another issue is to measure or implement the complexity of agents.


\end{document}